\documentclass[conference]{IEEEtran}
\IEEEoverridecommandlockouts
\usepackage{cite}
\usepackage{amsmath,amssymb,amsfonts}
\usepackage{algorithmic}
\usepackage{graphicx}
\usepackage{textcomp}
\usepackage{xcolor}
\def\BibTeX{{\rm B\kern-.05em{\sc i\kern-.025em b}\kern-.08em
    T\kern-.1667em\lower.7ex\hbox{E}\kern-.125emX}}

\usepackage{comment}
\usepackage{makecell}
\usepackage{booktabs} 
\usepackage{subcaption}
\usepackage{multirow}
\usepackage{tabularx}
\usepackage{float}
\usepackage{url}

\begin{document}

%==================================================================================
% Title
%==================================================================================
\title{AudioICL-Bench: A Benchmark for Large Audio Language Model In-Context Learning}

%==================================================================================
% Authors & Affiliations
%==================================================================================
\author{
\IEEEauthorblockN{Jia-Hung Chen}
\IEEEauthorblockA{\textit{National Taiwan University}\\
Taiwan \\
r14941279@ntu.edu.tw}
\and
\IEEEauthorblockN{Yi-Cheng Lin}
\IEEEauthorblockA{\textit{National Taiwan University}\\
Taiwan}
\and
\IEEEauthorblockN{Kai-Wei Chang}
\IEEEauthorblockA{\textit{Massachusetts Institute of Technology}\\
USA}
\and
\IEEEauthorblockN{Ke-Han Lu}
\IEEEauthorblockA{\textit{National Taiwan University}\\
Taiwan}
\and
\IEEEauthorblockN{Hung-Yi Lee}
\IEEEauthorblockA{\textit{NTU AI-CoRE}\\
Taiwan}
}
%\author{\IEEEauthorblockN{Anonymous Submission}}
\maketitle

% \begin{abstract}
%     In-context learning (ICL) lets models adapt to new tasks from a few demonstrations without parameter updates, an appealing prospect for audio, where labeling every new condition is costly.
%     Yet existing studies of audio ICL largely measure Task Recognition, where demonstrations merely cue a capability already acquired in pre-training, rather than Task Learning, where a model must infer a genuinely new input-label mapping from the demonstrations alone.
%     To test whether current Large Audio Language Models (LALMs) are capable of Task Learning, we introduce AudioICL-Bench, a diagnostic benchmark whose per-episode rules are resampled so that no correct answer can be recovered from prior knowledge.
%     Its nine tasks are organized along two axes, separating what must be learned from the demonstrations from what must be perceived in the signal, so that failures can be attributed to one or the other.
%     Across five LALMs, models readily bind arbitrary sounds to new labels when perception is easy, but collapse on temporal measurement, fine-grained acoustic discrimination, and composing several induced rules at once.
%     Crucially, these failures prove largely perceptual rather than learning failures, showing that the main bottleneck for audio Task Learning lies in perception itself.
% \end{abstract}

\begin{abstract}

    In-context learning (ICL) promises training-free adaptation for audio, where labeling every new condition is costly. Yet existing audio ICL studies largely measure Task Recognition, where demonstrations merely cue pre-trained capabilities, rather than Task Learning, where a genuinely new input–label mapping must be inferred from demonstrations alone. We introduce AudioICL-Bench, a diagnostic benchmark whose per-episode rules are resampled so that no correct answer is recoverable from prior knowledge. Its nine tasks are organized along two axes that separate what must be learned from demonstrations from what must be perceived in the signal, enabling failures to be attributed to either source. Across five Large Audio Language Models, the strongest models readily bind arbitrary sounds to new labels when perception is easy, but collapse on temporal measurement and composing multiple induced rules, revealing two primary capability boundaries: temporal perception and multi-rule composition. Relevant resources can be found at \textit{\url{https://github.com/robert0518/AudioICL-Bench}}. 

\end{abstract}

\section{Introduction}

Large Language Models (LLMs) can solve new tasks from only a few demonstrations without any parameter updates, an ability known as in-context learning (ICL)~\cite{brown2020language, min2022rethinking, von2023transformers, dong2024survey, si2023measuringinductivebiasesincontext}.
Recent Large Audio Language Models (LALMs)~\cite{cui2025recent, aroralandscape, wu2024towards, su2025audio, qwen3, qwenaudio} extend this paradigm to the acoustic domain, raising the prospect of training-free adaptation to new audio tasks where collecting labeled data or fine-tuning for every new environment is costly.
This prospect has motivated a growing body of work exploring ICL in audio and speech settings.
For example, Roll et al.~\cite{phi4} show that providing an LALM with speaker-matched utterances at inference time reduces ASR word error rates, and Piao et al.~\cite{alice} evaluate six LALMs on audio understanding tasks under progressively reduced textual guidance.
Other studies explore meta-training strategies that teach speech models to follow an ICL prompt format~\cite{textless,smile}, or leverage acoustically similar demonstrations to improve recognition in low-resource settings~\cite{li2026multimodalincontextlearningasr,zheng2025ticlcasestudyspeech,Li_2025}.
These results confirm that LALMs can benefit from in-context demonstrations, yet they raise a deeper question: does such improvement reflect genuine learning of new rules from the demonstrations, or does it merely reflect the model recognizing a familiar task and applying knowledge already acquired during pre-training?

\begin{figure}[t] 
    \centering
    \includegraphics[width=\columnwidth]{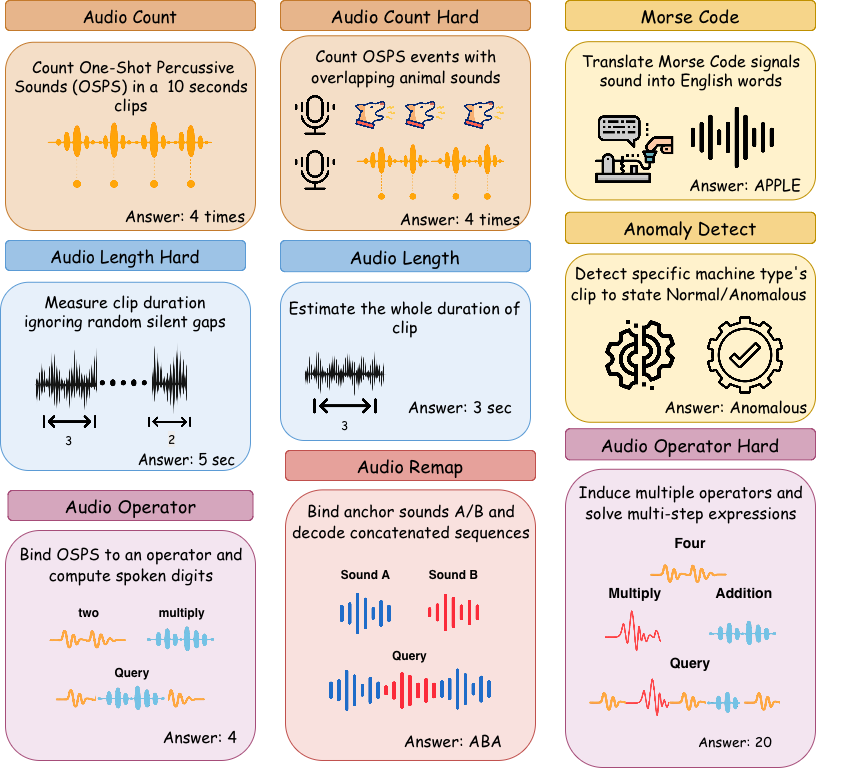}
    \caption{\textbf{AudioICL-Bench overview.} Each task is presented as an in-context episode: a few-shot sequence of audio--label demonstrations followed by a query whose label the model must predict. Episode-specific rules are resampled per episode, so the correct answer is recoverable only from the demonstrations and not from pre-trained priors.}
    \label{fig:audioicl-bench-overview}
\end{figure}

Pan et al.~\cite{pan-etal-2023-context} offer a framework that disentangles two mechanisms behind ICL.
\textbf{Task Recognition} captures the extent to which a model can recognize a task through demonstrations and apply its pre-trained priors; it can succeed even without ground-truth labels, because the demonstrations mainly cue an existing capability.
\textbf{Task Learning}, by contrast, is the ability to capture new input-label mappings that were unseen during pre-training, requiring the model to infer the underlying rule from the demonstrations and apply it to new queries.
The two mechanisms also exhibit distinct model size scaling behaviors: Task Recognition saturates quickly, whereas Task Learning emerges with model scale and improves steadily as more demonstrations are provided~\cite{pan-etal-2023-context}.

This distinction is especially consequential for audio, where many practical problems are defined by context-specific conventions that pre-trained knowledge cannot resolve.
A technician may demonstrate what counts as an abnormal machine sound on a specific factory line, or a field researcher may define a new bird-call category through a handful of labeled recordings; in both cases, the relevant rule must be inferred entirely from the demonstrations, placing such problems squarely in the regime of Task Learning.
Audio further compounds the challenge: unlike text, which is composed of discrete tokens with relatively stable semantic units, audio signals lack explicit unit boundaries, and the informative patterns may span multiple time scales, from sub-second timbral cues to multi-second event structures~\cite{suied2014auditory, mesaros2021sound}.
Models must therefore segment and interpret the signal perceptually before any rule can be induced from the demonstrations, making Task Learning in audio a qualitatively more demanding problem than in text.

Viewed through the lens of this framework, the audio ICL successes surveyed above are largely consistent with Task Recognition.
The tasks evaluated in these studies, including ASR, speech emotion recognition, gender recognition, spoken command classification, and language identification, are all well-established objectives whose label spaces and decision boundaries are fixed by the dataset and extensively encountered during pre-training~\cite{textless,smile,phi4,alice,li2026multimodalincontextlearningasr,zheng2025ticlcasestudyspeech,Li_2025}.
In each case, the demonstrations primarily convey contextual information about the test condition, such as speaker characteristics, accent, or domain, enabling the model to better apply an existing mapping rather than discover a new one.
To our knowledge, no existing audio ICL benchmark is specifically designed to isolate Task Learning, that is, to test whether LALMs can infer new rules from demonstrations when those rules cannot be recovered from pre-training alone.

To fill this gap, we introduce AudioICL-Bench, a diagnostic benchmark that isolates Task Learning in audio by building each task around an episode-specific rule that is resampled per evaluation instance.
We organize nine tasks along two complementary axes, a context-operation axis (Axis-C) and an audio-competency axis (Axis-A), defined in Section~\ref{sec:taxonomy}, and evaluate five representative LALMs.
Our main contributions are as follows:
\begin{itemize}
    \item \textbf{A diagnostic benchmark for audio Task Learning:} All tasks are built around episode-specific rules that are resampled per episode, ensuring that any correct prediction is  attributable to rule induction from the demonstrations rather than pre-trained priors.
    \item \textbf{A two-axis diagnostic taxonomy:} The context-operation axis captures what the model must learn or compose from demonstrations; the audio-competency axis captures what perceptual evidence must be extracted. Together they enable fine-grained diagnosis of where and why models fail.
    \item \textbf{Empirical analysis of five LALMs:} Models can perform symbolic binding when perceptual demands are modest, but consistently fail at temporal measurement and at composing multiple induced rules within a single episode, revealing clear capability boundaries.
\end{itemize}

%==================================================================================
% PART 1: Updated dataset.tex
%==================================================================================
 
\begin{table}[t] 
  \centering
  \caption{Overview of tasks in AudioICL-Bench. Source abbreviations: One-Shot Percussive Sounds (OSPS), Free Spoken Digit Dataset (FSDD), Wikimedia Commons (WC), Anomalous Sound Detection (ASD). Axis-C: C1\,=\,Fast Binding, C2\,=\,Induction, C3\,=\,Structured Composition. Axis-A: A1\,=\,Speech Content, A2\,=\,Non-Speech Acoustic Pattern, A3\,=\,Temporal Structure.}
  \label{tab:dataset}
  \small
  \fontsize{9}{10.5}\selectfont
  \setlength{\tabcolsep}{4pt} 
  \setlength{\aboverulesep}{0.4ex}
  \setlength{\belowrulesep}{0.4ex}
  \begin{tabular}{l l c c}
    \toprule
    \textbf{Task} & \textbf{Source} & \textbf{Axis-C} & \textbf{Axis-A} \\
    \midrule
    AudioCount      & OSPS \cite{oneshot} & C2 & A2 \\ 
    \midrule
    AudioCountHard  & \makecell[l]{OSPS \cite{oneshot} \\ ESC-50 \cite{ESC}} & C2 & A2 \\
    \midrule
    AudioLength     & Artificial & C2 & A3 \\
    \midrule
    AudioLengthHard & Artificial & C2 & A3 \\
    \midrule
    AudioOperator   & \makecell[l]{OSPS \cite{oneshot} \\ FSDD \cite{fsdd}} & C1, C2 & A1, A2 \\
    \midrule
    AudioOperatorHard & \makecell[l]{OSPS \cite{oneshot} \\ FSDD \cite{fsdd}} & C1, C2, C3 & A1, A2 \\
    \midrule
    AudioRemap      & OSPS \cite{oneshot} & C1, C3 & A2 \\
    \midrule
    MorseCode       & WC & C1, C3 & A3 \\
    \midrule
    AnomalyDetect   & ASD \cite{AD1, AD2} & C1 & A2 \\
    \bottomrule
  \end{tabular}
\end{table}

\section{AudioICL-Bench}
 
\subsection{Design and Taxonomy}\label{sec:taxonomy}
 
Table~\ref{tab:dataset} summarizes the nine tasks in AudioICL-Bench, and Figure~\ref{fig:audioicl-bench-overview} illustrates representative episodes.
As a diagnostic benchmark, AudioICL-Bench aims to isolate Task Learning from Task Recognition in the audio domain, and this diagnostic objective shapes two core design decisions.
 
First, all tasks are designed to be zero-shot resistant: without demonstrations, model performance remains near chance level. (One instruction-driven exception is analyzed in Section~\ref{sec:zero-shot}).
We achieve this by decoupling acoustic signals from their pre-trained semantic associations, for example by assigning arbitrary labels to sounds or requiring the induction of rules absent from pre-training corpora.
Any significant performance gain is therefore attributable to rule induction from the provided demonstrations.
 
Second, the benchmark primarily consists of artificial tasks, supplemented by one natural task.
Here, the artificial/natural distinction refers to task design rather than recording quality: an artificial task poses a scenario that would not arise in everyday practice (e.g., decoding Morse code, computing arithmetic from sound-encoded operators), whereas a natural task mirrors a scenario that practitioners actually face.
This follows a tradition of using artificial tasks to isolate specific capabilities under controlled conditions: Weston et al.~\cite{weston2015towards} pioneered synthetic tasks for diagnosing reasoning skills, Lake and Baroni~\cite{lake2018scan} showed that independently manipulating each task variable enables precise identification of where and why models fail, and Wei et al.~\cite{wei2023larger} demonstrated that decoupling labels from pre-trained semantic associations isolates genuine input-label learning from reliance on semantic priors.
AudioICL-Bench follows this principle: artificial task designs let us independently vary the context-side operation and the audio-side competency, which is difficult with natural tasks where these factors tend to co-vary.

Each task is organized into in-context \emph{episodes}.
An episode consists of a sequence of audio-label demonstrations followed by a query; the rule governing the correct answer is specific to that episode and resampled across episodes, so that pre-trained knowledge alone cannot produce correct predictions.
All tasks are evaluated using exact-match accuracy across 300 episodes per task.
Both the code and dataset will be publicly released to support reproducibility and future extension.
 
To systematically diagnose LALM capabilities, we organize tasks along two complementary axes.
The \textbf{context-operation axis (Axis-C)} characterizes what the model must learn or compose from the in-context demonstrations:
 
\begin{itemize}
    \item \textbf{C1. Fast Binding:} The demonstrations bind audio evidence to episode-specific labels or criteria, forming a temporary decision function. The general task interface is clear from the prompt, but the specific mapping (e.g., which sound maps to which label, or what counts as normal vs.\ anomalous) is defined only within the episode.
 
    \item \textbf{C2. Induction:} The demonstrations serve as the primary source for inferring a latent relation between audio evidence and outputs. The model must discover what aspect of the audio is relevant and what operation maps that evidence to the answer.
 
    \item \textbf{C3. Structured Composition:} The query requires structured, repeated, or multi-step application of bound or induced rules over multiple rule instances or extended sequences.
\end{itemize}
 
The \textbf{audio-competency axis (Axis-A)} characterizes what perceptual evidence must be extracted from the signal:
 
\begin{itemize}
    \item \textbf{A1. Speech Content Understanding:} Extracting linguistic or lexical content from speech, such as spoken digits or words.
 
    \item \textbf{A2. Non-Speech Acoustic Pattern Understanding:} Recognizing, comparing, or discriminating non-speech acoustic patterns. The required granularity ranges from broad event-level distinctions across different sound source families (e.g., dog bark vs.\ siren, percussive hit vs.\ alarm) to subtle within-domain variation where candidate sounds share a common source or operating context (e.g., normal vs.\ anomalous states of the same machine type~\cite{AD1, AD2}, or fault subtypes within vehicle engines~\cite{enginefault}).
 
    \item \textbf{A3. Temporal Structure:} Extracting temporal information such as event duration, onset/offset structure, or temporal code patterns (e.g., dot vs.\ dash timing in Morse telegraphy).
\end{itemize}
 
\noindent These two axes are orthogonal and compositional: a single task may require multiple labels on each axis, as shown in Table~\ref{tab:dataset}.
Together, they allow us to diagnose not only whether a model can perform Task Learning, but also which combination of context-side demand and audio-side competency causes failure.
In the current benchmark, the six A2 tasks and the three A3 tasks form two non-overlapping groups, providing a clean contrast between sound-event recognition and temporal measurement as potential sources of failure.
Within each group, variation in Axis-C labels enables further diagnosis of whether the bottleneck lies in binding, induction, or composition.

\begin{table*}[t]
\centering
\fontsize{9}{10.5}\selectfont
\setlength{\tabcolsep}{3.5pt}
\renewcommand{\arraystretch}{0.88}
\caption{Detailed evaluation of Qwen3-Omni-30B-A3B-Instruct on AudioICL-Bench across demonstration shots, with 95\% bootstrap CI. \textbf{All values are in \%}. Best in \textbf{Bold}. For AC, ACH, AL, and ALH, Specific $=$ General by construction (Section~\ref{sec:prompting}).}
\label{tab:detailed_results_qwen3_1col}
\begin{tabular}{l c *{9}{c}}
\toprule
\textbf{Strategy} & \textbf{k} & \textbf{AC} & \textbf{ACH} & \textbf{AL} & \textbf{ALH} & \textbf{AO} & \textbf{AOH} & \textbf{AR} & \textbf{MC} & \textbf{AD} \\
\midrule
\multirow{4}{*}{Specific}
& 0 & 0.3{$\pm$0.5} & 0.7{$\pm$0.8} & 0.0{$\pm$0.0} & 1.7{$\pm$1.5} & 34.0{$\pm$5.3} & 1.3{$\pm$1.2} & 56.0{$\pm$5.7} & 0.0{$\pm$0.0} & 49.7{$\pm$5.7} \\
& 1 & 43.0{$\pm$5.7} & 16.3{$\pm$4.2} & \textbf{8.0}{$\pm$3.2} & 11.0{$\pm$3.5} & 91.7{$\pm$3.2} & 4.7{$\pm$2.3} & 66.0{$\pm$5.3} & 0.0{$\pm$0.0} & 50.3{$\pm$5.7} \\
& 3 & 59.7{$\pm$5.7} & 21.0{$\pm$4.7} & 7.7{$\pm$3.0} & \textbf{16.0}{$\pm$4.0} & \textbf{99.7}{$\pm$0.5} & 5.3{$\pm$2.5} & 75.3{$\pm$4.8} & 0.0{$\pm$0.0} & 52.7{$\pm$5.7} \\
& 5 & \textbf{68.3}{$\pm$5.3} & \textbf{29.7}{$\pm$5.2} & \textbf{8.0}{$\pm$3.0} & 13.3{$\pm$3.8} & 98.7{$\pm$1.2} & \textbf{10.0}{$\pm$3.5} & \textbf{89.0}{$\pm$3.5} & 0.0{$\pm$0.0} & 55.3{$\pm$5.7} \\
\cmidrule(lr){1-11}
\multirow{4}{*}{General}
& 0 & 0.3{$\pm$0.5} & 0.7{$\pm$0.8} & 0.0{$\pm$0.0} & 1.7{$\pm$1.5} & 2.3{$\pm$1.8} & 2.3{$\pm$1.7} & 0.0{$\pm$0.0} & 0.0{$\pm$0.0} & 6.0{$\pm$2.8} \\
& 1 & 43.0{$\pm$5.7} & 16.3{$\pm$4.2} & \textbf{8.0}{$\pm$3.2} & 11.0{$\pm$3.5} & 73.0{$\pm$5.0} & 4.3{$\pm$2.2} & 35.0{$\pm$5.5} & 0.0{$\pm$0.0} & 47.7{$\pm$5.7} \\
& 3 & 59.7{$\pm$5.7} & 21.0{$\pm$4.7} & 7.7{$\pm$3.0} & \textbf{16.0}{$\pm$4.0} & 91.7{$\pm$3.2} & 4.3{$\pm$2.2} & 61.3{$\pm$5.5} & 0.0{$\pm$0.0} & 50.0{$\pm$5.7} \\
& 5 & \textbf{68.3}{$\pm$5.3} & \textbf{29.7}{$\pm$5.2} & \textbf{8.0}{$\pm$3.0} & 13.3{$\pm$3.8} & 96.7{$\pm$2.0} & 6.0{$\pm$2.7} & 82.7{$\pm$4.2} & 0.0{$\pm$0.0} & 54.7{$\pm$5.7} \\
\cmidrule(lr){1-11}
\multirow{3}{*}{Null}
& 1 & 0.0{$\pm$0.0} & 1.7{$\pm$1.5} & 5.3{$\pm$2.5} & 7.0{$\pm$2.8} & 20.7{$\pm$4.7} & 5.0{$\pm$2.5} & 18.7{$\pm$4.5} & 0.0{$\pm$0.0} & 50.0{$\pm$5.7} \\
& 3 & 1.0{$\pm$1.2} & 10.3{$\pm$3.5} & 7.0{$\pm$2.8} & 11.3{$\pm$3.7} & 64.0{$\pm$5.3} & 4.3{$\pm$2.2} & 43.7{$\pm$5.7} & 0.0{$\pm$0.0} & 52.3{$\pm$5.7} \\
& 5 & 5.0{$\pm$2.5} & 17.7{$\pm$4.3} & 7.0{$\pm$2.8} & 12.7{$\pm$3.8} & 73.7{$\pm$5.0} & 4.0{$\pm$2.2} & 62.3{$\pm$5.5} & 0.0{$\pm$0.0} & \textbf{56.7}{$\pm$5.7} \\
\bottomrule
\end{tabular}
\end{table*}
\begin{table*}[t]
\centering
\fontsize{9}{10.5}\selectfont
\setlength{\tabcolsep}{3pt}
\renewcommand{\arraystretch}{0.92}
\caption{Cross-model comparison at $k{=}5$ under Specific prompting (\%, with 95\% bootstrap CI). For AC, ACH, AL, and ALH the Specific prompt is the General prompt by construction (Section~\ref{sec:prompting}). A2 and A3 columns are the means over the six A2 tasks and three A3 tasks, respectively. Best per task in \textbf{bold}.}
\label{tab:cross_model}
\newcolumntype{Y}{>{\centering\arraybackslash}X}
\begin{tabularx}{\textwidth}{l *{6}{Y} | *{3}{Y} | c c}
\toprule
 & \multicolumn{6}{c|}{\textbf{A2 Tasks}} & \multicolumn{3}{c|}{\textbf{A3 Tasks}} & & \\
\textbf{Model} & \textbf{AC} & \textbf{ACH} & \textbf{AO} & \textbf{AOH} & \textbf{AR} & \textbf{AD} & \textbf{AL} & \textbf{ALH} & \textbf{MC} & \textbf{A2} & \textbf{A3} \\
\midrule
DeSTA   & 20.7{\tiny$\pm$4.7} & 28.3{\tiny$\pm$5.0} & 3.7{\tiny$\pm$2.2}  & 2.0{\tiny$\pm$1.5}           & 0.0           & 48.7{\tiny$\pm$5.7} & 3.3{\tiny$\pm$2.0} & \textbf{14.0}{\tiny$\pm$3.8}          & 0.0 & 17.2 & 5.8 \\
Qwen2   & 1.3{\tiny$\pm$1.2}  & 4.3{\tiny$\pm$2.3}  & 32.7{\tiny$\pm$5.3} & 2.3{\tiny$\pm$1.8}           & 9.3{\tiny$\pm$3.3}  & 43.3{\tiny$\pm$5.7} & 5.0{\tiny$\pm$2.5} & 8.3{\tiny$\pm$3.2}           & 0.0 & 15.6 & 4.4 \\
Qwen2.5 & 35.3{\tiny$\pm$5.3} & 26.0{\tiny$\pm$5.0} & 83.7{\tiny$\pm$4.2} & 2.3{\tiny$\pm$1.8}           & 18.0{\tiny$\pm$4.3} & 50.0{\tiny$\pm$5.7} & 5.3{\tiny$\pm$2.5} & 4.0{\tiny$\pm$2.2}           & 0.0 & 35.9 & 3.1 \\
Qwen3   & \textbf{68.3}{\tiny$\pm$5.3} & \textbf{29.7}{\tiny$\pm$5.2} & \textbf{98.7}{\tiny$\pm$1.2} & 10.0{\tiny$\pm$3.5} & \textbf{89.0}{\tiny$\pm$3.5} & \textbf{55.3}{\tiny$\pm$5.7} & \textbf{8.0}{\tiny$\pm$3.0} & 13.3{\tiny$\pm$3.8} & 0.0 & \textbf{58.5} & \textbf{7.1} \\
Gemini  & 36.0{\tiny$\pm$5.3} & 26.0{\tiny$\pm$4.8} & 77.7{\tiny$\pm$4.7} & \textbf{14.0}{\tiny$\pm$3.8} & 10.3{\tiny$\pm$3.5} & 51.0{\tiny$\pm$5.7} & 7.0{\tiny$\pm$2.8} & 7.3{\tiny$\pm$2.8}           & 0.0 & 35.8 & 4.8 \\
\bottomrule
\end{tabularx}
\end{table*}

\subsection{Artificial Tasks}

\noindent \textbf{AudioCount} (AC): The model is presented with an audio clip containing 1 to 8 non-overlapping instances of a percussive sound drawn from One-Shot Percussive Sounds (OSPS), placed on a silent background. The model must infer from the demonstrations that the label corresponds to the number of target events and apply this counting rule to the query. Chance accuracy is $1/8{\approx}12.5\%$.
 
\noindent \textbf{AudioCountHard} (ACH): This variant introduces acoustic interference by overlaying animal vocalizations from the ESC-50 dataset onto the target percussive events. The model must selectively attend to the target sound class amidst distractors while still performing event counting. The count range matches AC (1 to 8), so chance accuracy is also $1/8{\approx}12.5\%$.
 
\noindent \textbf{AudioLength} (AL): The model receives white noise segments ranging from 1 to 20 seconds. The demonstrations pair each segment with its duration as the label, and the model must infer this duration-estimation rule and apply it to the query. With integer-second labels over this range, chance accuracy is $1/20{=}5.0\%$.
 
\noindent \textbf{AudioLengthHard} (ALH): This variant inserts random periods of silence into an audio clip, splitting the audible content into multiple active segments. The model must identify the onset and offset of each audible segment and measure the total duration of all active audio. The active-region duration follows the same range as AL, so chance accuracy is also $1/20{=}5.0\%$.
 
\noindent \textbf{AudioOperator} (AO): Each query consists of a triplet: two spoken digits from the Free Spoken Digit Dataset (FSDD), interleaved by a percussive OSPS sound that represents one of three arithmetic operators (addition, subtraction, or multiplication). The sound-to-operator mapping is randomly drawn per episode. The model must bind each sound to its corresponding arithmetic function and compute the result. Since each operator yields a distinct numeric result, chance accuracy is $1/3{\approx}33.3\%$.
 
\noindent \textbf{AudioOperatorHard} (AOH): This extension uses multi-step arithmetic expressions (e.g., $n_1 \; \text{OSPS}_a \; n_2 \; \text{OSPS}_b \; n_3$), requiring the model to induce multiple sound-to-operator mappings within a single episode and compose them sequentially to produce the final answer. With two independent operator slots, chance accuracy is $(1/3)^2{\approx}11.1\%$.
 
\noindent \textbf{AudioRemap} (AR): In each episode, two anchor sounds ($A$ and $B$) drawn from OSPS are paired with symbolic labels that are given directly alongside each clip. The query and each demonstration present labeled $A$ and $B$ reference clips together with an unlabeled composite clip formed by concatenating three $A$/$B$ segments in some order. The model must match each segment in the composite clip to the correct anchor by acoustic similarity and output the corresponding label sequence in order. Chance accuracy for a length-3 sequence is $(1/2)^3{=}12.5\%$.
 
\noindent \textbf{MorseCode} (MC): The demonstrations provide an acoustic dictionary mapping each of the 26 English letters to its Morse code signal, along with example pairs of English words and their corresponding Morse audio sequences. The model must bind each letter to its acoustic pattern and infer the temporal conventions of Morse telegraphy (dot vs.\ dash timing, inter-character gaps) to decode novel Morse sequences into text. Because the output is a full word, chance accuracy is $(1/26)^L$ for a word of length $L$ and is not a single well-defined number across episodes; the single-character ablation in Section~\ref{sec:fail} isolates chance accuracy at $1/26{\approx}3.8\%$.

\subsection{Natural Task}\label{sec:natural-task}
 
\noindent \textbf{AnomalyDetect} (AD): Each episode is restricted to a single machine type drawn from the Anomalous Sound Detection dataset~\cite{AD1, AD2}. The demonstrations provide recordings labeled as \textit{Normal} or \textit{Anomalous} from that machine type, and the query is drawn from the same type. Unlike other A2 tasks in this benchmark, where candidate sounds differ at the level of broad event identity, AnomalyDetect requires within-domain discrimination: normal and anomalous recordings share the same source and spectral profile, and the model must detect subtle deviations from the demonstrated normal operating pattern. Chance accuracy is $1/2{=}50\%$.

\subsection{Prompting Strategy}\label{sec:prompting}

To disentangle the role of textual guidance from in-context rule induction, we evaluate models under three prompting configurations that vary the amount of task description while keeping the few-shot template fixed.
All configurations share the same input structure: a textual instruction (which may be empty) followed by $k$ demonstration slots and one query slot.
Each demonstration slot pairs one or more audio waveforms with a text label.
The general template for a single-clip task is:
\vspace{1pt}

\begin{quote}
\ttfamily \footnotesize
\textnormal{[Instruction]}\\[4pt]
{[Demo 1]} <|AUDIO|> Label: \textit{ground-truth}\\
\vdots\\
{[Demo \textit{k}]} <|AUDIO|> Label: \textit{ground-truth}\\
{[Query]} <|AUDIO|> Label:
\end{quote}

\noindent Here, bracketed markers such as \texttt{[Demo~\textit{i}]} and \texttt{[Query]} are literal text tags that appear in the prompt; \texttt{<|AUDIO|>} denotes an audio waveform passed through the model's native audio input channel and \texttt{Label:} is followed by the ground-truth label as text in demonstrations, but left blank in the query for the model to predict.
For multi-clip tasks such as AudioRemap, each demonstration slot contains multiple tagged audio clips (e.g., \texttt{[Demo~1~A]}, \texttt{[Demo~1~B]}, \texttt{[Demo~1~MIX]}), each followed by its own label.
For MorseCode, 26 reference entries mapping each letter to its Morse signal are prepended before the demonstrations.

The three configurations differ only in the \texttt{[Instruction]} field:

\begin{itemize}
    \item \textbf{Specific Prompting:} Each task receives a task-specific instruction describing its objectives. However, maintaining zero-shot resistance imposes a constraint on how specific these instructions can be: for tasks whose core operation must be inferred from demonstrations (AC, ACH, AL, ALH), any instruction naming the operation (e.g., ``counting,'' ``duration'') would resolve the task without requiring rule induction. We therefore use the General instruction as the Specific prompt for these four tasks by construction. This restricted-instruction design follows the principle adopted by VL-ICL Bench~\cite{vlicl}, which similarly avoids naming the target operation in tasks designed to test rule induction. For the remaining five tasks (AO, AOH, AR, MC, AD), the Specific instruction names the task-level objective without revealing the episode-specific mapping.

    \item \textbf{General Prompting:} All tasks receive a single uniform instruction: \textit{``Identify the hidden patterns and rules within the following audio-label pairs and apply them to the final query.''} This configuration tests whether models can induce rules without any task-specific priors.

    \item \textbf{Null Prompting:} The instruction field is left empty. The model receives only the raw few-shot demonstrations followed by the query, testing whether the audio-label structure alone is sufficient to trigger in-context learning.
\end{itemize}

\subsection{Evaluation Protocol}

For reproducibility, all models are decoded with temperature 0.0 and a maximum output length of 64 tokens.

Predictions are scored by deterministic parsing followed by exact matching, with task-specific extraction rules:
for integer-label tasks (AC, ACH, AO, AOH, AL, ALH), we extract the last integer from the model's response and compare it to the ground truth;
for AnomalyDetect, we perform keyword matching that normalizes common variants to \texttt{Anomalous} and maps the response to a binary label;
for AudioRemap, we extract alphabetic characters from the response, verify that the resulting sequence matches the expected length, and compare the uppercase sequence to the ground truth;
and for MorseCode, we apply normalized string matching after stripping whitespace and punctuation.

Zero-shot performance is reported in Table~\ref{tab:detailed_results_qwen3_1col} to contextualize prompt sensitivity; when a task has a finite output set, zero-shot accuracy is interpreted relative to its chance level.
Full prompts, evaluation code, and the dataset will be released upon publication.

\section{Results}

\begin{figure}[htbp]
    \centering
    \includegraphics[width= \columnwidth]{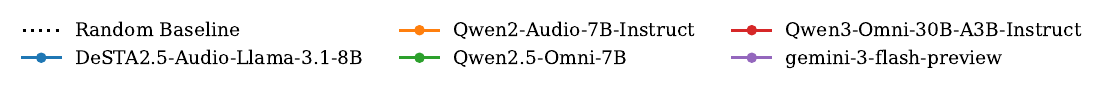} \\
    \vspace{0.3cm}
    \includegraphics[width= \columnwidth]{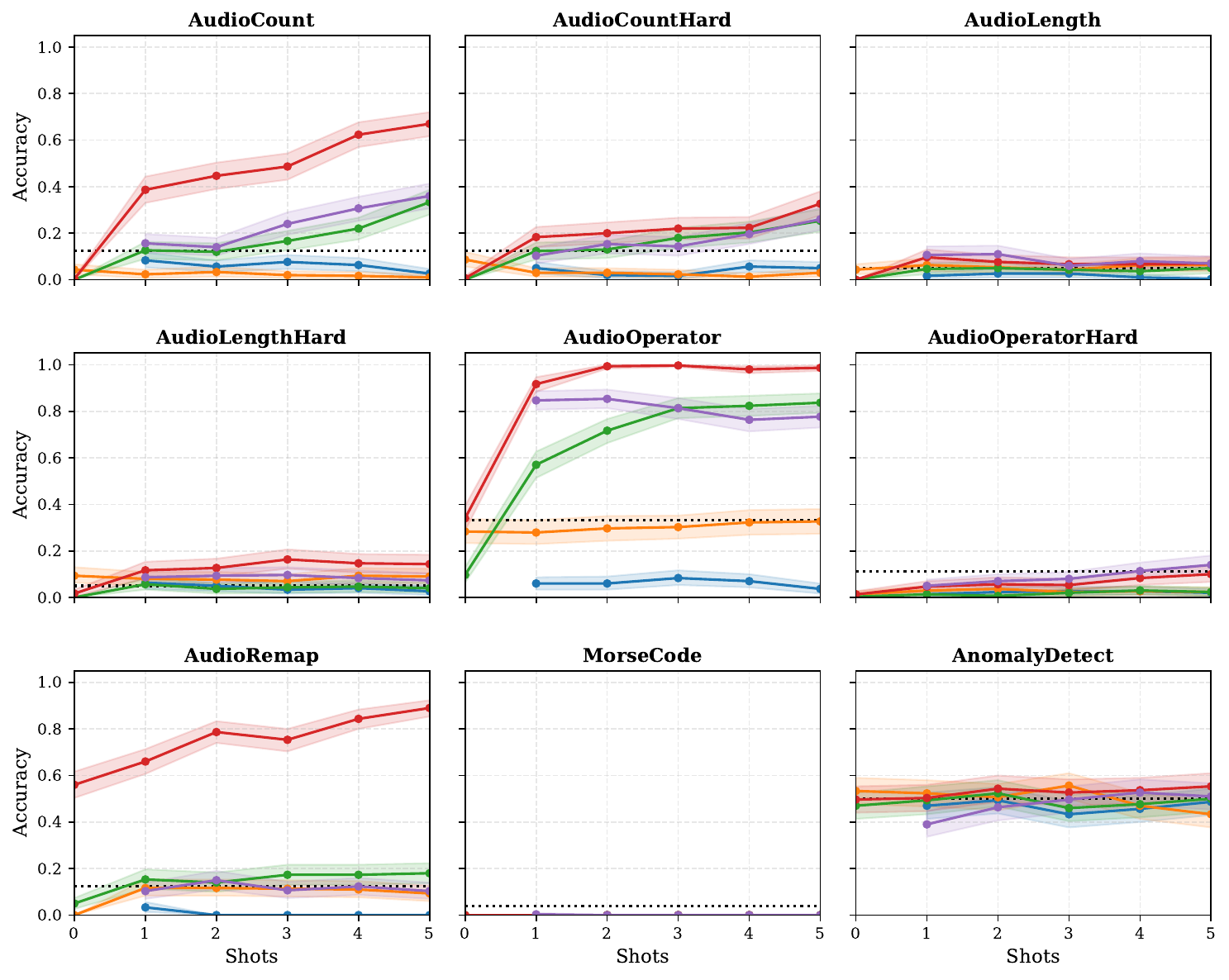}
    \caption{\textbf{Shot-scaling behavior across nine audio tasks under Specific prompting.} The $x$-axis is the number of in-context demonstrations $k$ (shots), and the $y$-axis is exact-match accuracy. Each colored line traces one of the five evaluated LALMs with shaded bands showing 95\% bootstrap confidence intervals; the dotted black line marks the chance-level baseline for each task.}
    \label{fig:shot_scaling_results}
\end{figure}

We evaluate five LALMs spanning different architectures and scales: DeSTA2.5-Audio-Llama-3.1-8B (DeSTA)~\cite{lu2025desta2}, Qwen2-Audio-7B-Instruct (Qwen2)~\cite{qwenaudio}, Qwen2.5-Omni-7B (Qwen2.5)~\cite{qwen25}, Qwen3-Omni-30B-A3B-Instruct (Qwen3)~\cite{qwen3}, and Gemini-3-flash-preview (Gemini).
Table~\ref{tab:detailed_results_qwen3_1col} reports results for Qwen3, the strongest model overall, across all three prompting configurations, and Figure~\ref{fig:shot_scaling_results} shows the shot-scaling curves under Specific prompting.

\subsection{Validating the Evaluation Design}

Before analyzing Task Learning capabilities, we first verify that our evaluation design successfully isolates Task Learning from Task Recognition by examining two conditions: zero-shot resistance and prompt sensitivity.

\noindent\textbf{Zero-shot resistance.}\label{sec:zero-shot}

At $k{=}0$ (no demonstrations), Qwen3 performs near or below chance on nearly all tasks, consistent with the design goal that pre-trained knowledge alone should not solve the tasks.
AnomalyDetect reaches 49.7\%, matching its 50\% chance baseline, and AudioOperator reaches 34.0\%, matching the $\sim$33\% chance of guessing one of three operators.
AudioRemap is an exception, reaching 56.0\% at $k{=}0$ under Specific prompting: because the anchor labels are given directly alongside the query clips, the task is solvable once the required matching operation and output format are understood, and the Specific instruction spells these out explicitly.
Under General prompting, which omits this operational detail, $k{=}0$ accuracy is 0.0\%, confirming that the instruction, not pre-trained knowledge, drives this exception.
These results confirm that pre-trained knowledge alone cannot solve the tasks, isolating Task Learning from Task Recognition.

\noindent\textbf{Prompt Sensitivity Affects Task Learning.}
As described in Section~\ref{sec:prompting}, the Specific prompt for AC, ACH, AL, and ALH is the General prompt by construction: any task-specific instruction would name the target operation and thereby resolve the task directly.
In contrast, removing all instructions (Null prompting) causes severe degradation on tasks that otherwise show learning, such as AudioCount (68.3\% vs.\ 5.0\% under Null) and AudioOperator (98.7\% vs.\ 73.7\%).
This confirms that textual instructions are often necessary to fully activate ICL for audio tasks, while the restricted instruction used for these four tasks does not leak the target rule.

\subsection{Where Does Audio Task Learning Succeed?}\label{sec:succeed}

Having established that observed performance reflects genuine Task Learning, we now use the two-axis taxonomy to characterize where this learning succeeds.
We focus on $k{=}5$ under Specific prompting, the setting that provides models with the most demonstrations while remaining within typical ICL budgets.

\noindent\textbf{Non-speech pattern recognition (A2) enables Task Learning.}
As shown in Table~\ref{tab:cross_model}, Qwen3 averages 58.5\% across the six A2 tasks at $k{=}5$.
The strongest results appear on AudioOperator (98.7\%) and AudioRemap (89.0\%), where the model successfully binds acoustic evidence to episode-specific labels or criteria and applies these bindings to novel queries.
AudioCount reaches 68.3\%, demonstrating that models can also induce an implicit counting rule from demonstrations alone.

\noindent\textbf{Shot-scaling confirms genuine Task Learning.}
On AudioCount (C2, A2), Qwen3 improves monotonically from 0.3\% at $k{=}0$ to 68.3\% at $k{=}5$ (Figure~\ref{fig:shot_scaling_results}).
This steady, shot-dependent improvement is a sign of Task Learning as characterized by Pan et al.~\cite{pan-etal-2023-context}, who showed that Task Learning, unlike Task Recognition, exhibits continuous improvement with additional demonstrations.
A similar pattern appears on AudioCountHard (0.7\% $\to$ 29.7\%) and AudioRemap (56.0\% $\to$ 89.0\%), while AudioOperator saturates rapidly (34.0\% $\to$ 99.7\% by $k{=}3$), consistent with the fast-binding nature of its C1 component.

\subsection{Where Does Audio Task Learning Fail?}\label{sec:fail}

\noindent\textbf{Temporal structure (A3) is a universal failure mode.}
In contrast to the A2 results, all three A3 tasks remain near zero across the entire model suite (Table~\ref{tab:cross_model}).
No model exceeds an A3-avg of 7.1\%, and the best single-task score is DeSTA's 14.0\% on AudioLengthHard.
Unlike A2 performance, which improves substantially across model generations (A2-avg: 15.6\% for Qwen2 $\to$ 58.5\% for Qwen3), A3 performance remains near zero regardless of model (A3-avg: 4.4\% $\to$ 7.1\%).

\begin{table}[htbp]
    \centering
    \small
    \renewcommand{\arraystretch}{0.9}
    \caption{Accuracy of single-character MorseCode inference}
    \label{tab:morse_single_char}
    \begin{tabular}{lc}
        \toprule
        \textbf{Model} & \textbf{Accuracy (\%)} \\
        \midrule
        Qwen2-Audio-7B-Instruct & 3.8 \\
        Qwen2.5-Omni-7B & 3.8 \\
        Qwen3-Omni-30B-A3B-Instruct & \textbf{11.5} \\
        \bottomrule
    \end{tabular}
\end{table}

To further isolate the source of failure, we conducted an ablation on MorseCode in which the 26-letter Morse dictionary is provided but only a single character is queried, completely removing the compositional demand (Table~\ref{tab:morse_single_char}).
Even in this reduced setting, Qwen3 achieves only 11.5\% accuracy, while the other two evaluated models collapse to the random-guess baseline of approximately 3.8\%.
This confirms that the bottleneck lies in temporal perception itself: the models cannot reliably distinguish the acoustic difference between a Morse dot and a Morse dash from the audio representation, even before any compositional decoding is required.

\subsection{Context-Side Complexity Interacts With Audio Competency}

The preceding analysis establishes that the audio-competency axis determines the primary capability boundary.
Within the A2 region where Task Learning is feasible, the context-operation axis (C-axis) reveals a secondary boundary that determines how complex the learned rules can be.

\noindent\textbf{Composition itself is not the bottleneck; composing multiple induced rules is.}

Three tasks jointly isolate where composition breaks down.
AudioRemap (C1$+$C3) requires composing bound mappings: each query segment must be matched to an anchor sound and the labels emitted in sequence. Qwen3 reaches 89.0\% at $k{=}5$, showing that structured, multi-step application of episode-specific mappings is itself within reach.
AudioOperator (C1$+$C2) instead requires inducing a sound-to-operator mapping, but applies only a single operation per query; Qwen3 reaches 99.7\% by $k{=}3$.
Only AudioOperatorHard (C1$+$C2$+$C3), which must induce multiple operator mappings within one episode and compose them sequentially, collapses to 10.0\% at $k{=}5$, a gap consistent across all models (Table~\ref{tab:cross_model}).
Since composition over bound mappings and single-rule induction each succeed in isolation, the bottleneck is their joint demand (C2$+$C3): composing multiple rules that must themselves be induced within a single episode, which exceeds current LALM capabilities even when the perceptual demands are manageable.

\noindent\textbf{Within-domain acoustic discrimination remains at chance.}
AnomalyDetect (C1, A2) poses the simplest context-side demand, fast binding of a normal/anomalous decision criterion, yet all models hover near the 50\% chance baseline (Table~\ref{tab:cross_model}).
As noted in Section~\ref{sec:natural-task}, this task requires within-domain discrimination where normal and anomalous recordings share the same source and spectral profile.
The uniform near-chance performance suggests that the acoustic difference between machine states is too subtle for current audio representations to capture from a few demonstrations, even when the context-operation demand is minimal.

\section{Conclusion}
We introduced AudioICL-Bench, a diagnostic benchmark that isolates Task Learning from Task Recognition in audio by requiring models to infer episode-specific rules from in-context demonstrations.
Organized along two complementary axes, audio competency and context operation, the benchmark reveals clear capability boundaries in current LALMs.
On the audio-competency axis, models achieve genuine Task Learning on non-speech pattern recognition tasks (A2), with the best model averaging 58.5\% across six A2 tasks, but universally fail on temporal structure tasks (A3), where no model exceeds 7.1\% on average.
The single-character Morse ablation localizes this failure to temporal perception itself, isolating it from the difficulty of sequential decoding.
On the context-operation axis, models handle single-step binding and induction (AudioOperator, 98.7\%) but fail when composition must operate over multiple rules induced within the same episode, as shown by the drop on AudioOperatorHard (10.0\%).
Within-domain acoustic discrimination also remains at chance, as evidenced by AnomalyDetect.
These findings suggest that advancing audio Task Learning will require representations that explicitly capture temporal structure and support compositional reasoning, capabilities that current audio encoders do not provide.

\section*{Acknowledgment}
This work was supported by the Ministry of Education (MOE) of Taiwan under the project Taiwan Centers of Excellence in Artificial Intelligence, through the NTU Artificial Intelligence Center of Research Excellence (NTU AI-CoRE)”.

\bibliographystyle{IEEEtran}
\bibliography{mybib}

\end{document}